%A SET OF MACROS FOR WRITING PAPERS.
%
%\today GIVES TODAY'S DATE.
%
\def\today{\ifcase\month\or January\or February\or March\or April\or May\or
June\or July\or August\or September\or October\or November\or December\fi
\space\number\day, \number\year}
%
%\note{footnote} GIVES SEQUENTIALLY NUMBERED FOOTNOTES.
%
\newcount\notenumber

\def\note{\global\advance\notenumber by 1 \footnote{$^{\the\notenumber}$}}
%
%\numbereq SEQUENTIALLY NUMBERS EQUATIONS ON THE RIGHT (number)
%
\newif\ifsectionnumbering
\newcount\eqnumber
\def\cleareqnumber{\eqnumber=0}
\def\numbereq{\global\advance\eqnumber by 1
\ifsectionnumbering\eqno(\the\secnumber.\the\eqnumber)
\else\eqno(\the\eqnumber) \fi}
\def\eqalinno{{\global\advance\eqnumber by 1}
\ifsectionnumbering(\the\secnumber.\the\eqnumber)\else(\the\eqnumber)\fi}
\def\name#1{\ifsectionnumbering\xdef#1{\the\secnumber.\the\eqnumber}
\else\xdef#1{\the\eqnumber}\fi}

\sectionnumberingtrue
%
%\ref{\name} GIVES SEQUENTIALLY NUMBERED REFERENCES [number], AND ASSIGNS
%THAT NUMBER TO A MACRO \name AND WRITES REF. TO FILE 1.
%
\newcount\refnumber

\immediate\openout1=refs.tex
\immediate\write1{\noexpand\frenchspacing}
\immediate\write1{\parskip=0pt}
\def\ref#1#2{\global\advance\refnumber by 1%
[\the\refnumber]\xdef#1{\the\refnumber}%
\immediate\write1{\noexpand\item{[#1]}#2}}
\def\tie{\noexpand~}

%
% NEW SECTION: \newsection The Method. (terminate with a .)
%
\font\twelvebf=cmbx10 scaled \magstep1
\newcount\secnumber

\def\newsection#1.{\ifsectionnumbering\cleareqnumber\else\fi%
	\global\advance\secnumber by 1%
	\bigbreak\bigskip\par%
	\line{\twelvebf \the\secnumber. #1.\hfil}\nobreak\medskip\par}
%
%
%\Box GIVES WAVE OPERATOR, OR LAPLACIAN
%
\def \sqr#1#2{{\vcenter{\vbox{\hrule height.#2pt
	\hbox{\vrule width.#2pt height#1pt \kern#1pt
		\vrule width.#2pt}
		\hrule height.#2pt}}}}

%
%
%\twocolumns GIVES TWO-COLUMN OUTPUT
%
\newdimen\fullhsize
\def\fiddle{\fullhsize=6.5truein \hsize=3.2truein}
\def\fullline{\hbox to\fullhsize}
\def\mkhdline{\vbox to 0pt{\vskip-22.5pt
	\fullline{\vbox to8.5pt{}\the\headline}\vss}\nointerlineskip}
\def\mkftline{\baselineskip=24pt\fullline{\the\footline}}
\let\lr=L \newbox\leftcolumn
\def\twocolumns{\fiddle
	\output={\if L\lr \global\setbox\leftcolumn=\columnbox
		\global\let\lr=R \else \doubleformat \global\let\lr=L\fi
		\ifnum\outputpenalty>-20000 \else\dosupereject\fi}}
\def\doubleformat{\shipout\vbox{\mkhdline
		\fullline{\box\leftcolumn\hfil\columnbox}
		\mkftline} \advancepageno}
\def\columnbox{\leftline{\pagebody}}
\magnification=1200
\def\pr#1 {Phys. Rev. {\bf D#1\tie }}
\def\pre#1 {Phys. Rep. {\bf #1\tie}}
\def\pe#1 {Phys. Rev. {\bf #1\tie}}
\def\pl#1 {Phys. Lett. {\bf #1B\tie }}
\def\prl#1 {Phys. Rev. Lett. {\bf #1\tie }}
\def\np#1 {Nucl. Phys. {\bf B#1\tie }}
\def\ap#1 {Ann. Phys. (NY) {\bf #1\tie }}
\def\cmp#1 {Commun. Math. Phys. {\bf #1\tie }}
\def\imp#1 {Int. Jour. Mod. Phys. {\bf A#1\tie }}
\def\mpl#1 {Mod. Phys. Lett. {\bf A#1\tie}}
%% poor man's black board bold
\def\BbbZ{{}\kern+1.6pt\hbox{$I$}\kern-7.5pt\hbox{$Z$}}

\def\tie{\noexpand~}
\def\ov{\overline}
\def\s{(\sigma)}

\def\sp{(\sigma ')}

\def\Tb{\overline T}
\def\d{\delta(\sigma-\sigma ')}
\def\dpr{\delta'(\sigma-\sigma ')}

\def\dpppr{\delta'''(\sigma-\sigma ')}

\def\dx{\partial X}
\def\dbx{{\overline\partial}X}
\def\dsx{\partial^2X}
\def\dbsx{{\overline\partial}^2X}
\parskip=15pt plus 4pt minus 3pt
\headline{\ifnum \pageno>1\it\hfil  T-Duality in Arbitrary
	$\ldots$\else \hfil\fi}
\font\title=cmbx10 scaled\magstep1
\font\tit=cmti10 scaled\magstep1
\footline{\ifnum \pageno>1 \hfil \folio \hfil \else
\hfil\fi}
\raggedbottom
\rightline{\vbox{\hbox{RU95-04}\hbox{CTP-TAMU-28/95}}}
\vfill
\centerline{\title T-DUALITY IN ARBITRARY}
\vskip 20pt
\centerline{\title STRING BACKGROUNDS}
\vfill
{\centerline{\title Mark Evans${}^{a,c}$
and
Ioannis Giannakis${}^{a,b}$ \footnote{$^{\dag}$}
{\rm e-mail: evans@theory.rockefeller.edu,
giannak@theory.rockefeller.edu}}
}
\medskip
\centerline{$^{(a)}${\tit Physics Department, Rockefeller
University}}
\centerline{\tit 1230 York Avenue, New York, NY
10021-6399}
\medskip
\centerline{$^{(b)}${\tit Center for Theoretical Physics,
Texas A{\&}M
University}}
\centerline{\tit College Station, TX 77843-4242}
\medskip
\centerline{$^{(c)}${\tit Citicorp Securities, Inc.}}
\centerline{\tit 399 Park Avenue., New York, NY 10017.}

\vfill
\centerline{\title Abstract}
\bigskip
{\narrower\narrower
T-Duality is a poorly understood symmetry of the space-time fields
of string theory that interchanges long and short distances. It is
best understood in the context of toroidal compactification where,
loosely speaking, radii of the torus are inverted.  Even in this
case, however, conventional techniques permit an understanding of
the transformations only in the case where the metric on the torus
is endowed with Abelian Killing symmetries. Attempting to apply
these techniques to a general metric appears to yield a non-local
world-sheet theory that would defy interpretation in terms of
space-time fields. However, there is now available a
simple but powerful general approach to understanding the symmetry
transformations of string theory, which are generated by certain
similarity transformations of the stress-tensors of the associated
conformal field theories. We apply this method to the particular
case of T-Duality and i) rederive the known transformations, ii)
prove that the problem of non-locality is illusory, iii) give an
explicit example of the transformation of a metric that lacks Killing
symmetries and iv) derive a simple transformation rule for arbitrary
string fields on tori.
\par}
\vfill\vfill\break

\newsection Introduction.

One of the severest tests for a theory of particle physics with any
claim to completeness is the manner in which it deals with the
ultra-violet problems of field theory. It would appear that string
theory does this in a uniquely appealing manner. Rather than
delicately adjusting itself to cancel the ultra-violet divergences,
string theory appears simply to abolish the ultra-violet---something
it can do because of the dynamical nature of space-time in a theory
of gravity. It is fair to say that this phenomenon is understood
only incompletely, but string theory, in a completely differentiable
way, seems to deny the existence of distances significantly shorter
than the Planck length. If we try to probe shorter and shorter
distances, a reinterpretation of the degrees of freedom shows that
we are deluding ourselves, and are, in fact, probing longer and
longer distances: the ultra-violet eludes our grasp by turning out
to be the infra-red in disguise. Such a fascinating phenomenon would
have many attractive consequences, but one in particular stands out
for its philosophical importance. For most theories of particle
physics, a skeptic may always wonder whether it is simply an effective
theory, derived from something more fundamental at shorter distances,
and whether this process of reduction continues indefinitely to
shorter and shorter distances. String theory, if it truly denies
the existence of shorter distances, can perhaps lay some claim to
finality.

Unfortunately, we understand this phenomenon only in examples. The
simplest is the one-loop diagram in perturbation theory. Here, we
are instructed to integrate over a single copy of the moduli-space
of the torus, the famous fundamental region, which is usually defined
in such a way that the modular parameter, $ \tau$, never approaches
the origin that, naively, would describe arbitrarily small loops.
More directly, gedanken-experiments at ultra-Planckian energies
\ref\gross{D.\tie
Gross, \prl60 (1988), 1229; D.\tie Gross and P.\tie Mende, \np303
(1988), 407; D.\tie Amati, M.\tie Ciafaloni and G.\tie Veneziano,
\pl216 (1989), 41.} show the need to adjust the uncertainty principle
in such a way that there is a maximal spatial resolution, no matter
how great the momentum.

However, the most studied example of this phenomenon is surely
T-Duality of compactified string theory (For a recent review see
\ref\secon{A.\tie Giveon, M.\tie Porrati and E.\tie Rabinovici,
\pre244 (1994).}). It is not  a totally new idea, being a strong-weak
coupling duality  of the world-sheet theory, of a type going back
to  Kramers and Wannier \ref\kramwan{H. A.\tie Kramers
and G. H.\tie Wannier, \pe60 (1941), 252 .}, but
it made its first appearance
in string theory in the work of Kikkawa and Yamasaki and of Sakai
and Senda \ref\first{K.\tie Kikkawa and M.\tie Yamasaki, \pl149
(1984), 357; N.\tie Sakai and I.\tie Senda, Prog. Theor. Phys. {\bf
75} (1986), 692.}. These authors considered a string moving on a
space-time in which one spatial dimension is compactified---a circle
of radius $R$ (in suitable units of the Planck length). In such a
background there are two noteworthy types of excitation:  strings
with (integer quantized) momenta in the compact direction, and
strings winding around the compact dimension an integer number of
times. The masses of the momentum excitations are of the form ${}\sim
n/R$, with integral $n$, while those of the winding modes are ${}\sim
m R$, where $m$ is the number of times the string wraps around the
compact dimension. Already, there is apparent a duality in which
interchanging the r\^{o}le of the momentum and winding modes is
equivalent to mapping $R\rightarrow 1/R$. This result was further
generalised in \ref{\vpn}{V. P. Nair, A. Shapere, A. Strominger and
F. Wilczek, \np287 (1987), 402.}.  Also suggestions have been made
in the literature towards an understanding of T-duality as a canonical
transformation  \ref{\eqwvio}{A. Giveon, E. Rabinovici and G.
Veneziano, \np322 (1989), 167.}, \ref\cann{K. A. Meissner and G.
Veneziano \pl267 (1991), 33; E. Alvarez, L. Alvarez-Gaume and Y.
Lozano, \pl336 (1994), 183.} and a duality symmetric formulation
of string world-sheet dynamics was proposed in
\ref{\tse}{A. A. Tseytlin, \pl242 (1990), 163; \np350 (1991), 395.}.

To see this duality in more detail, consider the simplest case of a
string moving on a circle. The world-sheet stress tensor is
$$
T(\sigma) = {\textstyle {1 \over 4}} R \left( \pi(\sigma)/R
+ X'(\sigma)\right)^2,
\numbereq\name{\eqstress}
$$
where $X(\sigma)$ is the coordinate of the
string (of period $2\pi$) and $\pi(\sigma)$ is its conjugate momentum:
$$
\left[\pi(\sigma), X'(\sigma') \right] = i \delta'(\sigma - \sigma')
\numbereq\name{\eqalg}
$$
However, $\pi(\sigma)$ and $X'(\sigma)$ are
algebraically indistinguishable in
equation (\eqalg), which means that we could just as
well interpret equations (\eqstress) and (\eqalg) as
describing a string with coordinate $\tilde X(\sigma)$
and momentum $\tilde\pi(\sigma)$ given by,
$$
\eqalign{\tilde\pi(\sigma) &= X'(\sigma) \cr
		\tilde X'(\sigma) &= \pi(\sigma). \cr}
\numbereq\name{\dualalg}
$$
In terms of {\it these} variables, the stress tensor is
$$
T(\sigma) = {\textstyle {1 \over 4}} R \left( \tilde X'(\sigma)/R
+ \tilde\pi(\sigma)\right)^2
		= {\textstyle {1 \over 4}} 1/R \left(R \tilde \pi(\sigma)
+ \tilde X'(\sigma)\right)^2,
\numbereq\name{\eqdualstress}
$$
which, comparing with equation (\eqstress), clearly
describes a string moving on a circle of radius $1/R$.

In essence, this is as far as the problem is
understood (we shall briefly mention various
embellishments below), but this understanding is only
partial. The algebra automorphism of equation (\dualalg) tells
us how to map $X'(\sigma)$, but is silent on the
transformation of $X(\sigma)$ itself, or, worse yet, would
seem to imply a transformation which is nonlocal on the world-sheet:
$$
\tilde X(\sigma)  \sim  \int^{\sigma} dx \pi(x).
\numbereq\name{\eqforgot}
$$

If we wished to make a T-Duality transformation on a configuration where
the stress-tensor depended on $X(\sigma)$ as well as $X'(\sigma)$, we
would apparently end up with an unacceptable (and uninterpretable)
non-local stress-tensor. This happens as soon as we consider, for
example, field configurations where the space-time metric is not flat
(We should mention that we shall not concern ourselves in this paper
with locality of the theory in space-time, our use of the word local,
will refer exclusively to the world-sheet. For work relevant to this issue
see \ref\nonl{I. Bakas, \pl343 (1995), 103; E. Alvarez,
L. Alvarez-Gaume and I. Bakas,
\np457 (1995), 3; S. F. Hassan, CERN preprint CERN-TH-95-290
(1995), hep-th/9511056; K. Kounnas, \pl321 (1994), 26; I. Antoniadis,
S. Ferrara and K. Kounnas, \np421 (1994), 343.}).

To the best of our knowledge, this problem has
been resolved only partially in the literature, and progress
in understanding T-Duality has been limited to finding as many circumstances
as possible where it need not be confronted.

The simplest such extension is to consider space-time compactified
on a product of several circles endowed with a flat metric. This
straightforward generalization of the one-circle case gives a
separate duality for each circle, and this approach was pushed to its limit
in the work of  Buscher \ref\buscher{T. Buscher, \pl 159 (1985), 127,
\pl 194 (1987), 59 and \pl 201 (1988), 466.}, who showed how to
implement T-duality transformations on manifolds with Abelian
isometries. In essence, the existence of $n$ Killing fields with
vanishing Lie-bracket for any pair means that a coordinate system
may be found in which the space-time metric is independent of $n$
coordinates. Thus it is not surprising that it is possible to implement
a separate T-Duality for each such cyclic coordinate. The results
for tori may also be extended to orbifolds, which are flat almost
everywhere \ref\orbrefs{ J.\tie Lauer, J.\tie Mas and H. P.\tie Nilles, \pl226
(1989), 251; J.\tie Lauer and R.\tie
Zucchini, \pl235 (1990), 268; R.\tie Zucchini, \np350 (1991), 111.
}, WZW and coset models \ref\kir{E.\tie Kiritsis, \np405 (1993), 109; \mpl6
(1991), 2871.}.
The technique most commonly used, gauging the isometries and
then getting the two dual descriptions of the same system with different
choices of fields to be integrated out, is not the one described above, but
the results and scope are the same \ref\gau{
M.\tie Rocek and E.\tie Verlinde, \np373 (1992), 630.}
. In particular, isometries may be gauged only if they exist, so this technique
does {\it not} permit us to dualize configurations which depend in an essential
way on all the coordinates. That world-sheet locality is preserved in these
cases was proven by Hassan \ref\hass{S. F. Hassan, CERN preprint
CERN-TH-95-98 (1995), hep-th/9504148.}, who showed that it
followed from conservation of the isometry current.
Aspects of non-abelian dualities were addressed in \ref\nab{X. De la Ossa and
F. Quevedo, \np403 (1993), 109; A. Giveon and M. Rocek, \np421 (1994), 173;
M. Gasperini, R. Ricci and G. Veneziano, \pl319 (1993),
438; E. Kiritsis and N. A. Obers,
\pl334 (1994), 67; C. Klimcik and P. Severa, \pl351 (1995), 455; E. Alvarez,
L. Alvarez-Gaume and Y. Lozano, \np424 (1994), 155.}.

How, then, are we to dualize string configurations where the
action or stress-tensor of the CFT depend on $X(\sigma)$ in an
essential way? Equation (\eqforgot) might lead us to think that the
problem is intractable or, worse yet, uninteresting. However, equation
(\eqforgot) is the result of an essentially {\it classical} world-sheet
analysis. Our main purpose in this paper is to explain how to do a
fully quantum mechanical analysis, and to demonstrate that, when
this is done, {\it all world-sheet non-localities are  cancelled!} We regard
this as a string miracle of the first rank.

The crucial observation was made some time ago by Dine, Huet and Seiberg
\ref\dhs{M.\tie Dine, P.\tie Huet and N.\tie Seiberg, \np 322 (1989), 301.},
who observed that T-duality is, in fact, a finite gauge transformation. The
Kaluza-Klein mechanism tells us that if a generally covariant theory (such as
string theory) is
compactified on a  circle, a $U(1)$ gauge symmetry will result. (For string
theory it is actually $U(1)\times U(1)$, because string theory possesses
two-form gauge invariance in addition to general covariance). It is a famous
result of string theory that this unbroken gauge symmetry is enhanced
to $SU(2)\times SU(2)$ when the radius of the circle has a critical value. The
observation of Dine, Huet and Seiberg is that applying these additional
gauge transformations with a parameter of the correct magnitude changes
the sign of the Kaluza-Klein dilaton, inverting the radius of
compactification. This observation seems more natural, although no
less profound, when we recall
that the extra gauge bosons are winding modes, while the Kaluza-Klein
excitations are momentum modes. The enhanced gauge symmetry therefore
mixes winding and momentum modes, just as T-Duality interchanges them.

In recent years our understanding of the gauge invariances
of string theory has improved considerably  \ref{\ovre}{M.\tie Evans and
B.\tie Ovrut, \pr41 (1990), 3149; \pl231 (1989), 80.}, \ref{\eaw}{M. Evans
and I. Giannakis, \pr44, 2467 (1991).}.
We understand (in principle, at least) how to implement gauge
transformations on arbitrary backgrounds. Combining this knowledge
with the DHS insight should therefore enable us to understand T-Duality
applied to arbitrary field configurations. As we shall see, this hope is
fully realized.

How are gauge symmetries implemented? The simple answer is
that a similarity transformation should be applied to the stress tensor
of the conformal field theories:
$$
T(\sigma) \longmapsto e^{ih} T(\sigma) e^{-ih}.
\numbereq\name\eqsim
$$
This yields a new stress tensor from which we may
read off the transformed space-time fields. (the stress tensor is to string
theory exactly what a superfield is to supergravity). What is the
operator $h$ that implements a gauge transformation? For each gauge
symmetry there exists a corresponding current algebra on the
world-sheet, generated by, say, $J^a(\sigma)$ ($a$ spans
the adjoint representation of the gauge algebra). For a gauge transformation
with parameter $\Lambda^a(X)$, the generator, $h$, in equation (\eqsim) is just
$$
h = \int\; d\sigma \> \Lambda^a(X(\sigma)) J^a(\sigma).
\numbereq\name\eqgen
$$
It really is that simple.

In the next section we shall explain this approach to the study of
gauge symmetry in string theory in more detail, summarizing earlier
work, while in section 3 we shall demonstrate the power and versatility
of these methods by deriving most of the known results on T-Duality.
Section 4 is devoted to a discussion of general backgrounds; we
shall dualize a simple non-flat background that lacks isometries,
and derive a general transformation rule for arbitrary background
fields. As advertized, it is local. Section 5 is devoted to some
concluding remarks.

\newsection Deformations of Conformal Field Theories and
Symmetries.

In this section we shall review earlier work  [\ovre], [\eaw] on
deformations of conformal field theories and symmetries of string
theory.  For more details the reader is referred to the original
papers, or the review contained in \ref\dgmtalk{M.\tie Evans and
I.\tie Giannakis in S.\tie Catto and A.\tie Rocha, eds., {\it
Proceedings of the XXth International Conference on Differential
Geometric Methods in Theoretical Physics}, World Scientific, Singapore
(1992), p.\tie 759, hep-th/9109055.}. That rare reader already
familiar with this work may skip this section without loss.

To study symmetries, we seek transformations of the space-time
fields that take one solution of the classical equations of motion
to another that is physically equivalent.  Since, ``Solutions of
the classical equations of motion," are, for the case of string
theory \ref{\soln}{C.\tie Lovelace, \pl135 (1984), 75; C.\tie Callan,
D.\tie Friedan, E.\tie Martinec and M.\tie Perry, \np262 (1985),
593; A.\tie Sen, \pr32 (1985), 2102.}, two-dimensional conformal
field theories \ref{\bpza}{A. Belavin, A. Polyakov and A. Zamolodchikov
\np241 (1984), 333; D. Friedan, E. Martinec and S. Shenker \np271
(1986), 93.}, we are thus interested in physically equivalent
conformal field theories.

Any quantum mechanical theory (including a CFT) is defined
by three elements: i)~an algebra
of observables, $\cal A$ (determined by the degrees of
freedom of the theory
and their equal-time commutation relations),  ii)~a
representation of that algebra
and iii)~a distinguished element of $\cal A$ that generates
temporal evolution
(the Hamiltonian). Note that for the same $\cal A$ we may
have many choices
of Hamiltonian, so that $\cal A$ should more properly be
associated with a
{\it deformation class\/} of theories than with one
particular theory.
For a CFT, we further want $\cal A$ to be generated by
local fields,
$\Phi(\sigma)$ (operator valued distributions on a circle
parameterized
by $\sigma$), and we require not just a single
distinguished operator, but
two distinguished fields, $T(\sigma)$ and $\ov T(\sigma)$.
In terms of these fields
the Hamiltonian, $H$, and generator of translations, $P$,
may be written
$$
\eqalignno{H&=\int d\sigma (T\s +
\Tb\s)&\eqalinno\name{\eqham}\cr
P&=\int d\sigma (T\s - \Tb\s)&\eqalinno\name{\eqmom}\cr}
$$
and they must satisfy Virasoro$\times$Virasoro:
$$
\eqalignno{[T\s, T\sp]&={- i c \over 24\pi}\dpppr
+2 i T\sp\dpr -  i T'\sp\d&{\global\advance\eqnumber by 1}
(\the\secnumber.\the\eqnumber a)\name{\eqvir}\cr
[\Tb\s,\Tb\sp]&={ i c \over 24\pi}\dpppr
-2 i\Tb\sp\dpr +  i\Tb'\sp\d&(\the\secnumber.\the\eqnumber
b)\cr
[T\s,\Tb\sp]&=0.&(\the\secnumber.\the\eqnumber c)\cr}
$$
Except on $\sigma$, a prime denotes differentiation.
$T$ and $\ov T$ are the non-vanishing components of the
stress-tensor,
and must satisfy (\eqvir) if they are to generate
conformal transformations.
Also of interest are the so-called
{\it primary fields\/} of dimension $(d,\ov d)$, $\Phi\s$,
defined
by the conditions
$$
\eqalign{[T\s,\Phi_{(d,\overline d)}\sp]&=i d
\Phi_{(d,\overline d)}\sp
\dpr-( i/\sqrt2)\partial\Phi_{(d,\overline d)}\sp\d\cr
[\Tb\s,\Phi_{(d,\overline d)}\sp]&=- i\overline d
\Phi_{(d,\overline d)}
\sp\dpr-( i/\sqrt2)\overline\partial\Phi_{(d,\overline
d)}\sp\d\cr}
\numbereq\name{\eqprim}
$$

Clearly, then, two CFTs will be physically equivalent if
there is an isomorphism
between the corresponding algebras of observables, $\cal
A$, that maps
stress tensor to stress tensor. (The mapping of primary to
primary is then
automatic).
The simplest example of such an isomorphism
is an inner automorphism, or similarity transformation
$$
\Phi(\sigma)\mapsto
e^{ih}\Phi(\sigma)e^{-ih},\numbereq\name{\eqinaut}
$$
for any fixed operator $h$. {\it Thus the physics will be
unchanged if we change
a CFT's stress tensor by just such a similarity
transformation.}

Now, the stress tensor is parameterized by the space-time
fields of  string theory. For example,
$$
T_{G_{\mu\nu}}(\sigma) = \textstyle {1\over2} G_{\mu\nu}(X)
\partial X^\mu\partial X^\nu
\numbereq\name{\eqgravback}
$$
corresponds to the space-time metric $G_{\mu\nu}$, with
all other fields
vanishing. Thus an appropriate similarity transformation (\eqinaut)
applied to $T$
will produce a change in $T$ which
corresponds to a change
in the space-time fields, {\it without changing the
physics}. This change
in the space-time fields is therefore a {\it symmetry\/}
transformation.

We may clarify
the way in which the change in the stress tensor may be
interpreted as a
change in the space-time fields by first considering the
more
general problem of deforming a conformal field theory (we
now consider
deformations which, while they preserve conformal
invariance,
{\it need not be symmetries, e.g.} we may deform flat empty space
so that a weak gravitational wave propagates through it).
It is straightforward to
show that, to first order, the Virasoro algebras (\eqvir)
are preserved by
deforming the choice of stress tensor by a so-called {\it
canonical
deformation} [\ovre],
$$
\delta T\s = \delta\Tb\s =
\Phi_{(1,1)}\s \numbereq\name{\eqcan}
$$
where $\Phi_{(1,1)}\s$ is a primary
field of dimension (1,1) with respect to the stress
tensor \footnote*{This is a straightforward consequence
of the definition of a primary field, eq. (\eqprim)
and the form of the Virasoro algebras, eq. (\eqvir).}. We reiterate:
(\eqcan) does {\it not\/} in general correspond to a
symmetry transformation,
although it preserves conformal invariance.
Since (1,1) primary
fields are vertex operators for physical states, they are
in natural
correspondence with the space-time fields, and equation
(\eqcan) makes the
connection between changes of the stress tensor and
changes of the space-time
fields more transparent.

Returning now to the problem of
symmetries, if we take
the generator $h$ in equation (\eqinaut) to be the zero
mode of an
infinitesimal (1,0) or (0,1) primary field (a current),
then it is
straightforward to see
that its action on the stress tensor is necessarily
a canonical deformation, as in equation (\eqcan),
and so may be translated easily into a change in the
space-time fields (for examples, see [\ovre]). It is
well known that {\it conserved\/} currents generate
symmetries \ref{\band}{T.\tie Banks and L.\tie Dixon, \np307
(1988), 93.}, but
within the formalism described here, conservation is {\it
not\/}
necessary, a fact that does not seem to have been widely
appreciated.
Indeed, it is not hard to see that a non-conserved current
generates
a symmetry that is spontaneously broken by the particular
background being
considered [\ovre].

At this point it is convenient to assess the strengths and
weaknesses of our analysis so far---canonical deformations
and symmetries generated by the zero-modes of currents. We
begin with the strengths:{\parskip=\smallskipamount
\item{$\bullet$} The fact that a current zero-mode generates a
canonical deformation guarantees that we can translate the
inner automorphism into a transformation on the physical
space-time fields.
\item{$\bullet$} In this way, one may exhibit symmetries both familiar
(general coordinate and two-form gauge transformations, regular
non-abelian gauge transformations---including the Green-Schwarz
modification---for the heterotic string) and unfamiliar
(an infinite class of spontaneously broken, level-mixing gauge
supersymmetries) [\ovre].\par}

\noindent However, there are also deficiencies:
{\parskip=\smallskipamount
\item{$\bullet$} Explicit calculations are hard to perform for a general
configuration of the space-time fields. It would appear that we need to
know the precise form of the stress-tensor and its currents for that
general configuration (recall that a current is only a current relative
to a particular stress-tensor). This is a very hard problem, essentially
requiring us to find the general solution to the equations of motion
before we can discuss symmetries.
\item{$\bullet$} It is hard to say anything about the symmetry algebra.
This is just the algebra of the generators, but it is {\it not\/} in
general true that the commutator of two current zero-modes is itself a
current zero-mode: our symmetry ``algebra" does not close!
\item{$\bullet$} Finally, by
considering a few examples, it is easy to see that the
canonical deformation of equation (\eqcan) corresponds to
turning on
space-time fields {\it in
a particular gauge\/} (something like Landau or harmonic
gauge), and so
symmetries generated by zero-modes of currents preserve
this gauge
condition, since they generate canonical deformations. We would like
to understand the gauge principle behind string theory without
imposing gauge conditions.\par}

It turns out that these three drawbacks are intimately related, and
may be largely overcome by moving beyond canonical deformations,
equation (\eqcan), and beyond the zero-modes of currents as symmetry
generators. Equation (\eqcan) is {\it not\/} the
most general infinitesimal deformation that preserves the
Virasoro
algebras (\eqvir). In [\eaw] we showed that, for the
massless degrees
of freedom of the bosonic string in flat space, we could
find a distinct
deformation of the stress tensor for each solution of the
linearized
Brans-Dicke equations. This correspondence was found by
considering the
general translation invariant {\it ansatz\/} of naive
dimension two for
$\delta T$;
$$
\eqalignno{
\delta T=&H^{\nu\lambda}(X)\dx_\nu\dbx_\lambda+
A^{\nu\lambda}(X)\dx_\nu\dx_\lambda+\cr
&\qquad\qquad B^{\nu\lambda}(X)\dbx_\nu\dbx_\lambda
+C^\nu(X)\dsx_\nu+D^\lambda(X)\dbsx_\lambda,&
\eqalinno\name{\eqansatz}\cr
}
$$
with a similar, totally independent {\it ansatz\/} for
$\delta\Tb$.
The fields
$H^{\mu\nu}$ {\it etc.}~turn out to be characterized in
terms of solutions
to the linearized Brans-Dicke equation when we demand that
the deformation
preserves (to first order) the Virasoro algebras (\eqvir).

By considering
this more general {\it ansatz}, we get more than just
covariant equations of
motion---we also understand a larger set of symmetry
generators, $h$.
Indeed, any generator that preserves the form of the {\it
ansatz\/}
(\eqansatz)
must necessarily generate a change in the stress tensor
that corresponds
to a change in the space-time fields.

The condition that $\delta T$ be of naive dimension two
(with which we
shall soon dispense) is preserved if $h$ is of naive
dimension zero. The
condition of translation invariance is
$$
\left[P,\delta T\s\right] = -i\delta T'\s,
\numbereq\name{\eqtransinv}
$$
which may be preserved by demanding that $h$ commute with
$P$, the
generator of translations, (\eqmom). (Equation (\eqtransinv)
may also be thought
of as a gauge condition, but not one that has any obvious
interpretation
in terms of the space-time fields). Taken together, these
conditions
characterize $h$ as the zero-mode of a field of naive
dimension one.

The lesson to be drawn from this massless example is
clear: the way to
introduce space-time fields unconstrained by gauge
conditions is to consider
an {\it arbitrary translation invariant ansatz\/} for the
deformation
of the stress tensor, and to ask only that it preserve the
Virasoro
algebras. To move beyond the massless level, we simply
drop the requirement
that the naive dimension be two. We argued in [\eaw] that, as
with the superfield
formulation of supersymmetric theories,
this was likely to
introduce auxiliary fields beyond the massless level, but so be it.
(Indeed, this
whole formulation
of string theory is completely parallel to a superspace approach,
with
$T$ and $\Tb$ as superfields and
derivatives of the world-sheet scalars playing the r\^ole
of the
odd coordinates of superspace).

Having dropped any
requirement on the naive dimension of $\delta T$, we know
that {\it any}
operator $h$ that commutes with $P$ will generate a
symmetry transformation
on our space-time fields (possibly including the
auxiliaries). This extension of the set of symmetry generators
ameliorates each of the drawbacks mentioned above:
{\parskip=\smallskipamount
\item{$\bullet$} Since the symmetry generators are no longer
currents and the stress-tensor is generic, we no longer need
explicit forms for the general CFT or its currents. To give explicit
symmetry transformations on our expanded set of fields we need only
calculate the commutator of the general zero-mode with the general
field. Even this is problematic, however; see \ref{\asios}{M. Evans,
I. Giannakis and D. Nanopoulos \pr50 (1994), 4022.}
\item{$\bullet$} While zero-modes of currents do not, in general,
close under commutation, the set of {\it all\/} zero-modes {\it
does\/} close. This algebra may be characterized as the centralizer
of $P$, the generator of world-sheet translations, equation (\eqmom).
\item{$\bullet$} It should also be emphasized that our symmetry algebra
is now {\it background independent}. Recall that, naively at least,
the algebra of observables, $\cal A$, is attached to a {\it deformation
class\/} of CFTs, the elements of which differ only in their choice of
stress-tensor. Our set of symmetry generators---all zero-modes---is
manifestly independent of any particular choice of stress-tensor, and
so is their algebra. (Formally, this follows from the fact that $P$
can deform by at most a central element, so that its centralizer is
invariant under deformation.)\par}

For example, we may generate general coordinate and two-form
gauge transformations (at least about flat space) by choosing
$$
h={\int}d{\sigma}[{\xi}^{\mu}(X){\partial}X_{\mu}+{\zeta}^{\mu}(X)
{\overline{\partial}}X_{\mu}]  \numbereq\name{\eqabouo}
$$
with ${\delta}T$ given by (\eqansatz). In the next section we shall
identify the operator $h$ which implements T-duality in string theory.

\newsection Duality in spaces with abelian isometries.

In the previous section we have sketched how symmetries of string
theory ({\it i.e.} symmetry transformations on the spacetime fields of string
theory) are generated by inner automorphisms acting on the operator
algebra of the theory. These inner automorphisms, {\it i.e.} just similarity
transformations, generate the infinite dimensional symmetry algebra
that underlies string theory. It includes both the unbroken symmetries
(space-time diffeomorphisms, gauge transformations)
and an infinite class of spontaneously gauge broken
(super) symmetries (higher symmetries, dualities) which mix
different mass levels. The symmetry generators (zero modes of
operators which implement the inner automorphisms) are independent
of any particular choice of stress-tensor since the algebra of
observables is attached to a {\it deformation class} of CFT's.
In this section we will identify the operator $h$ which maps the
operator algebra onto itself and in addition can be pulled back to
spacetime and be interpreted as a T-duality transformation on the
spacetime fields. This will be achieved by fixing the operator
algebra and constructing the operator $h$ at the self-dual point.
The effect of a T-duality transformation on space-time fields can
be calculated then by applying the inner automorphism to an
arbitrary stress-tensor in the {\it deformation class}.

For illustrative purposes we will initially consider a simple
example, string propagation on a circle
of radius $R$. Since the coordinate of the string
$X(\sigma, \tau)$ parametrizes a circle $S^1$, it obeys a periodicity
condition
$$
X({\sigma}+2{\pi}, \tau)=X(\sigma, \tau)+2{\pi}nR.
\numbereq\name{\eqroiu}
$$
The most general expression for $X(\sigma, \tau)$
satisfying the two-dimensional wave equation and consistent
with the boundary condition, Eq. (\eqroiu) then becomes
$$
X(\sigma, \tau)=x+p{\tau}+nR{\sigma}+i{\sum_{n \ne 0}}{1\over n}
({a_n}e^{-in({\tau}-{\sigma})}+{{\tilde a}_n}e^{-in({\tau}+{\sigma})}).
\numbereq\name{\eqfoud}
$$
The periodicity condition of the string coordinate $X(\sigma, \tau)$
implies that the momentum $p$ is quantized, $p={m\over R}$,
while the second term in the boundary condition of $X(\sigma, \tau)$
describes winding string states. The integer $n$ counts how many times
the string wraps around the circle.
The two components of the world-sheet stress-tensor which describe
this particular conformal field theory
are given by
$$
T_R(\sigma)={\textstyle{1\over 2}}:\hat{\partial X}\hat{\partial X}:(\sigma)
\qquad
{\overline T}_R(\sigma)={\textstyle{1\over 2}}:\hat{{\overline{\partial}}X}
\hat{\overline{\partial}}X:(\sigma)
\numbereq\name{\eqfuxo}
$$
where we have defined the stress-tensor through a point
splitting regularization as follows:
$$
T_R(\sigma)={\textstyle{1\over 2}}:\hat{\partial X}\hat{\partial X}:(\sigma)
={\lim_{{\epsilon}\to 0}}
{\textstyle{1\over 2}}\hat{\partial X}(\sigma)\hat{\partial X}(\sigma+\epsilon)
+{\textstyle{1\over 4{\pi}{\epsilon}^2}}
\numbereq\name{\eqdoye}
$$
The subscript $R$ indicates that the stress-tensor depends on the
radius of compactification $R$ which, since
it may vary, becomes a spacetime field.

We should now address a minor technical point that can nonetheless
be very confusing. The operator written as $\hat{\partial X}(\sigma)$ in
equation
(\eqdoye) is not the same operator at different radii. This is apparent if
we express it in terms of $\pi(\sigma)$ and $X(\sigma)$:
$$
\hat{\partial X}={1\over {\sqrt 2}}({{\sqrt {\alpha}'}\over R}{\pi}
+{R\over {\sqrt {\alpha}'}}X') \qquad
\hat{{\overline{\partial}}X}={1\over {\sqrt 2}}
({{\sqrt {\alpha}'}\over R}{\pi}-{R\over {\sqrt {\alpha'}}}X')
\numbereq\name{\eqdxoi}
$$
$\pi(\sigma)$ and $X(\sigma)$ have fixed, $R$-independent, commutation
relations,
$$
[\pi(\sigma), X(\sigma')]=i{\delta}({\sigma}-{\sigma'})
\numbereq\name{\eqfasoul}
$$
while the operators  $\hat{\partial X}(\sigma)$ do not.
If we want to compare CFT's at different values of $R$, it is essential that
we express the stress-tensors in terms of fixed, $R$-independent
operators, such as $\pi(\sigma)$ and $X(\sigma)$. It is slightly more
convenient, however, to work instead with the light-cone derivatives at the
critical radius. Thus, by the symbols ${\partial X}$ and
${{\overline{\partial}}X}$ (without the hat), we shall mean
$$
{\partial X}={1\over {\sqrt 2}}({{\sqrt {\alpha}'}\over R_{cr}}
{\pi}(\sigma)+{R_{cr}\over {\sqrt {\alpha'}}}X'(\sigma)) \qquad
{{\overline{\partial}}X}={1\over {\sqrt 2}}
({{\sqrt {\alpha}'}\over R_{cr}}
{\pi}(\sigma)-{R_{cr}\over {\sqrt {\alpha'}}}X'(\sigma)),
\numbereq\name{\eqdxoi}
$$
neither more nor less. At the critical radius these are indeed the
light-cone derivatives,
but for other values of $R$ they are not. Since we shall never
actually be interested in taking light-cone derivatives,
there is no danger of ambiguity.

We saw in equation (\dualalg) that $T$-duality involves the interchange of
$\pi(\sigma)$ and $X'(\sigma)$, and so we seek an operator
$h$ that achieves this; we need
$$
e^{i{\pi}h}{{\sqrt {\alpha}'}\over R_{cr}}{\pi}(\sigma)e^{-i{\pi}h}=
-{R_{cr}\over {\sqrt {\alpha'}}}X'(\sigma) \qquad e^{i{\pi}h}
{R_{cr}\over {\sqrt {\alpha'}}}X'(\sigma)
e^{-i{\pi}h}=-{{\sqrt {\alpha}'}\over R_{cr}}{\pi}(\sigma),
\numbereq\name{\eqcoua}
$$
which from the definition (\eqdxoi) is equivalent to,
$$
e^{i{\pi}h}{\partial X}(\sigma)
e^{-i{\pi}h}=-{\partial X}(\sigma) \qquad e^{i{\pi}h}
{\overline{\partial}}X(\sigma)e^{-i{\pi}h}={\overline{\partial}}X(\sigma).
\numbereq\name{\eqcima}
$$

To find this operator, $h$, we draw on the famous result that, at
the critical radius, {\it i.e.} $R=R_{cr}=\sqrt 2$, the $U(1)_L \times
U(1)_R$ gauge symmetry of the theory extends to $SU(2)_L \times SU(2)_R$.
This symmetry enhancement is due to the appearance of extra $(1,0)$ and
$(0,1)$ operators $e^{{\pm i}{\sqrt 2}X_L}(\sigma)$, $e^{{\pm i}
{\sqrt 2}X_R}(\sigma)$. Subsequently the operators (${\sqrt 2}i{\partial}X
(\sigma)$,
$e^{{\pm i}{\sqrt 2}X_L}(\sigma)$) form an $SU(2)_L$ current algebra.
It is straightforward now to construct the operator $h$ if we recall
the familiar formula for the generators of $SU(2)$
$$
e^{i{\pi}J_2}J_3e^{-i{\pi}J_2}=-J_3
\numbereq\name{\eqvolm}
$$
as follows
$$
h={1\over 2i}{\int}d{\sigma}(e^{i{\sqrt 2}X_L}-e^{-i{\sqrt 2}X_L})(\sigma)
\numbereq\name{\eqxiou}
$$
We can verify that this particular choice of $h$ satisfies
Eq. (\eqcima) by writing
$$
e^{i{\pi}h}{\partial X}(\sigma)
e^{-i{\pi}h}
={\partial X}(\sigma)
+i{\pi}{\lbrack h, {\partial X}(\sigma) \rbrack}
+{1\over 2}(i{\pi})^2{\lbrack h, {\lbrack h, {\partial X}(\sigma) \rbrack},
\rbrack}+\cdots
\numbereq\name{\eqasopu}
$$
and calculating the commutators explicitly. This construction of $h$ is
simply a consequence of combining the insight of Dine, Seiberg
and Huet [\dhs], that $T$-duality is an enhanced gauge transformation, with
our understanding that gauge transformations are implemented
through inner automorphisms generated by the corresponding
world-sheet current algebra, as explained in equation (\eqgen).

Having constructed the operator $h$ which implements the inner automorphism
of the operator algebra, we proceed to calculate its effect on the
stress-tensor of the theory, Eq. (\eqdoye),
at  a general value of $R$. As was explained above,
we must first express the stress tensor in terms of an $R$-independent
basis of the operator algebra. Thus we need to
express ${\hat{\partial X}}(\sigma)={1\over {\sqrt 2}}
({{\sqrt {\alpha'}}\over
R}{\pi}(\sigma)+{R\over {\sqrt {\alpha'}}}X'(\sigma))$
in terms of ${\partial X}(\sigma)
={1\over {\sqrt 2}}({{\sqrt {\alpha}'}\over R_{cr}}{\pi}
(\sigma)+{R_{cr}\over {\sqrt {\alpha'}}}X'(\sigma))$
and ${\overline{\partial}}X(\sigma)={1\over {\sqrt 2}}
({{\sqrt {\alpha}'}\over R_{cr}}{\pi}(\sigma)
-{R_{cr}\over {\sqrt {\alpha'}}}X'(\sigma))$,
$$
{\hat{\partial X}}(\sigma)={1\over 2}[{{R_{cr}^2}\over
R}({\partial X}+{\overline{\partial}}X)+R({\partial X}-{\overline
{\partial}}X)](\sigma)
\numbereq\name{\eqduax}
$$
Substituting into Eq. (\eqdoye) we obtain the following expression for
the stress-tensor at radius $R$
$$
\eqalign{
T_R={\lim_{{\epsilon}\to 0}}
{1\over 8}&{\lbrack ({{R_{cr}}^2\over R}+R)^2{\partial X}(\sigma){\partial X}
({\sigma}+{\epsilon})+({{R_{cr}}^2\over R}-R)^2
{\overline{\partial}}X(\sigma)
{\overline\partial}X({\sigma}+{\epsilon})} \cr
&+{({{R_{cr}}^4\over R^2}-R^2)
({\partial X}(\sigma){\overline{\partial}}X({\sigma}+{\epsilon})+
{\overline{\partial}}X(\sigma){\partial X}({\sigma}+{\epsilon}))+{1\over
4{\pi}{\epsilon}^2} \rbrack} \cr}
\numbereq\name{\eqscop}
$$
Since the inner automorphism generated by $h$ changes the sign
of $\partial X$, its effect on $T_R$ is simply obtained:
$$
\eqalign{
e^{i{\pi}h}{T_R}(\sigma)&e^{-i{\pi}h}={\lim_{{\epsilon}\to 0}}
{1\over 8}{\lbrack ({{R_{cr}}^2\over R}+R)^2{\partial X}(\sigma){\partial X}
({\sigma}+{\epsilon})+({{R_{{cr}}}^2\over R}-R)^2
{\overline{\partial}}X(\sigma)
{\overline{\partial}}X({\sigma}+{\epsilon})} \cr
&+{(-{{R_{cr}}^4\over R^2}+R^2)
({\partial X}(\sigma){\overline{\partial}}X({\sigma}+{\epsilon})+
{\overline{\partial}}X(\sigma){\partial X}({\sigma}+{\epsilon}))+{1\over
4{\pi}{\epsilon}^2} \rbrack} \cr
&={\lim_{{\epsilon}\to 0}}
{1\over 8}{\lbrack ({{R_{cr}}^2\over {\tilde R}}
+{\tilde R})^2{\partial X}(\sigma){\partial X}
({\sigma}+{\epsilon})+({{R_{{cr}}}^2\over {\tilde R}}-{\tilde R})^2
{\overline{\partial}}X(\sigma)
{\overline{\partial}}X({\sigma}+{\epsilon})} \cr
&+{({{R_{cr}}^4\over {\tilde R}^2}-{\tilde R}^2)
({\partial X}(\sigma){\overline{\partial}}X({\sigma}+{\epsilon})+
{\overline{\partial}}X(\sigma){\partial X}({\sigma}+{\epsilon}))+{1\over
4{\pi}{\epsilon}^2} \rbrack}=T_{{\tilde R}={R_{cr}^2\over R}}(\sigma) \cr}
\numbereq\name{\eqscop}
$$
Thus the inner automorphism generated by $h$
maps the world-sheet stress-tensor onto a different one.
The resulting conformal field theory is isomorphic to the original one.
But this particular automorphism can be interpreted as
a transformation on the spacetime field $R$, the radius of the circle.
Hence the transformation $R\to {\tilde R}={R_{cr}^2\over R}$
is a symmetry of string theory.

Next we turn our attention to a more generalized setting: string
propagation on a $D$-dimensional flat torus in the presence of
a constant antisymmetric background field $b_{\mu\nu}$. In one of
the symmetric points of the {\rm deformation class}
the gauge symmetry $[U(1)_L]^D
\times [U(1)_R]^D$ of a generic point is enhanced
to $[SU(2)_L]^D \times [SU(2)_R]^D$. The stress-tensor of
this particular conformal field theory, represented by the symmetric
point under consideration, is given by
$$
T_G(\sigma)={1\over 2}G^{\mu\nu}:{\partial}
X_{\mu}{\partial}X_{\nu}:(\sigma)
\qquad {\overline T}_G(\sigma)={1\over 2}G^{\mu\nu}:{\overline{\partial}}
X_{\mu}{\overline{\partial}}X_{\nu}:(\sigma)
\numbereq\name{\eqsoaz}
$$
where $G_{\mu\nu}$ is a constant diagonal metric (the identity!) and the
antisymmetric background field has been set to zero.
As in the previous example, the symbols ${\partial}X_{\mu}(\sigma)$
and ${\overline{\partial}}X_{\mu}(\sigma)$ should be thought of
merely as the combinations
$$
{\partial}X_{\mu}(\sigma)={1\over {\sqrt 2}}({\pi}_{\mu}
+G_{\mu\nu}X'^{\nu})(\sigma), \qquad {\overline{\partial}}X_{\mu}(\sigma)=
{1\over {\sqrt 2}}({\pi}_{\mu}-G_{\mu\nu}X'^{\nu})(\sigma)
\numbereq\name{\eqszpa}
$$
There are $D$ independent inner automorphisms in this case, one for each
dimension of the torus. Since at the symmetric point the
symmetry group is enhanced to $[SU(2)_L]^D \times [SU(2)_R]^D$, they
are generated by the $J_2$ generators of
the several $SU(2)$'s.
The corresponding operators $h^{(i)}$ which
implement the inner automorphisms
are thus given by
$$
h^{(i)}={1\over 2i}{\int}d{\sigma}(e^{i{\sqrt 2}k^{(i)}_\mu X^{\mu}}-
e^{-i{\sqrt 2}k^{(i)}_\mu X^{\mu}}),
\numbereq\name{\eqsoaxi}
$$
where $k^{(i)}_\mu$ is a suitable basis of Killing forms on the
torus. For a $D$-dimensional flat torus, there are $D$ of them $(i
=1, \cdots, D)$ and we have chosen a particular basis where $k^{(i)}=
(1,0,0, \cdots), (0,1,0, \cdots), (0,0,1, \cdots), \ldots $.

As we deform our conformal field theory away from the
symmetric point the stress-tensor
changes and at a generic point of the {\rm deformation class} takes
the form
$$
T_{g,b}(\sigma)={1\over 2}g^{\mu\nu}:{\hat{\partial
X_{\mu}}}{\hat{\partial X_{\nu}}}:(\sigma)
\qquad {\overline T}_{g,b}(\sigma)={1\over 2}g^{\mu\nu}:
\hat{{\overline\partial}
X_{\mu}}\hat{{\overline\partial}X_{\nu}}:(\sigma),
\numbereq\name{\eqsytu}
$$
where
$$
{\hat{\partial X_{\mu}}}(\sigma)={1\over {\sqrt 2}}({\pi}_{\mu}
+(g_{\mu\nu}+b_{\mu\nu})X'^{\nu})(\sigma), \qquad
\hat{{\overline\partial}X_{\mu}}(\sigma)=
{1\over {\sqrt 2}}({\pi}_{\mu}-(g_{\mu\nu}+b_{\mu\nu})X'^{\nu})(\sigma).
\numbereq\name{\eqsmz}
$$
Note that $g$ represents a generic (constant) metric, while $G$ is the
metric at the point of enhanced symmetry.

As in our previous simple example, we first must
express ${\hat{\partial X_{\mu}}}$ in
terms of ${\partial X_{\mu}}$ and ${\overline{\partial}}X_{\mu}$
as follows:
$$
{\hat{\partial X_{\mu}}}={1\over 2} {\lbrack ({\partial X_{\mu}}+
{\overline{\partial}}X_{\mu})+(g_{\mu\rho}+b_{\mu\rho})G^{\rho\nu}
({\partial X_{\nu}}-{\overline{\partial}}X_{\nu}) \rbrack}
\numbereq\name{\eqauios}
$$
Substituting into Eq. (\eqsytu) we get
$$
\eqalign{
&T_{g,b}(\sigma)={1\over 8} {\lbrack
(g^{\mu\nu}+g^{\mu\rho}(g_{\rho\sigma}
+b_{\rho\sigma})G^{\sigma\nu}+g^{\rho\nu}(g_{\rho\sigma}
+b_{\rho\sigma})G^{\mu\sigma}} +g^{\rho\sigma}(g_{\rho\kappa}
+b_{\rho\kappa})(g_{\sigma\lambda}+b_{\sigma\lambda})\cr
&G^{\mu\kappa}G^{\lambda\nu}){\partial X_{\mu}}{\partial X_{\nu}}
+(g^{\mu\nu}-g^{\mu\rho}(g_{\rho\sigma}
+b_{\rho\sigma})G^{\sigma\nu}-g^{\rho\nu}(g_{\rho\sigma}
+b_{\rho\sigma})G^{\mu\sigma}+g^{\rho\sigma}(g_{\rho\kappa}
+b_{\rho\kappa})\cr
&(g_{\sigma\lambda}+b_{\sigma\lambda})
G^{\mu\kappa}G^{\lambda\nu}){\overline{\partial}}X_{\mu}
{\overline{\partial}}X_{\nu}
+{(g^{\mu\nu}-g^{\rho\sigma}(g_{\rho\kappa}
+b_{\rho\kappa})(g_{\sigma\lambda}+b_{\sigma\lambda})
G^{\mu\kappa}G^{\lambda\nu})
({\partial X_{(\mu}}{\overline{\partial}}X_{\nu)}) \rbrack} \cr}
\numbereq\name{\eqcerou}
$$
As we remarked above, there are now $D$ separate $T$-dualities. We shall
consider them separately below, but for now we shall follow the historical
route, and consider just the product of all $D$ of them. The corresponding
transformation on the stress tensor is, therefore,
$$
e^{i{\pi}h^{(1)}} \cdots e^{i{\pi}h^{(n)}}
T_{g,b}e^{-i{\pi}h^{(1)}}\cdots e^{-i{\pi}h^{(n)}}(\sigma)=T_{{\tilde g},
{\tilde b}} ,
\numbereq\name{\eqcerq}
$$
where ${\tilde g}^{\alpha\beta}$ and ${\tilde b}^{\nu\sigma}$ satisfy
$$
{\tilde g}^{\alpha\beta}=g^{\mu\nu}(g_{\mu\rho}+b_{\mu\rho})(g_{\nu\sigma}
+b_{\nu\sigma})G^{\rho\alpha}G^{\sigma\beta} \qquad
{\tilde g}^{\alpha\nu}({\tilde g}_{\nu\sigma}+{\tilde b}_{\nu\sigma})=
g^{\beta\nu}(g_{\nu\rho}+b_{\nu\rho})G^{\rho\alpha}G_{\sigma\beta}.
\numbereq\name{\equrew}
$$
These two relations can be summarized as
$$
{\tilde g}^{\mu\nu}+{\tilde b}^{\mu\nu}=(g_{\kappa\lambda}
+b_{\kappa\lambda})G^{\kappa\mu}G^{\lambda\nu}.
\numbereq\name{\eqasopi}
$$

The separate T-duality transformations generated by the
individual $h^{(i)}$, Eq. (\eqsoaxi), are part of
the $O(d, d, Z)$ group of dualities of toroidal compactifications
discovered in [\eqwvio], and
correspond to the so-called factorized dualities. The remaining $O(d, d, Z)$
duality transformations are implemented by different inner automorphisms of
the operator algebra \ref{\innwr}{I. Giannakis, Rockefeller
University preprint-RU04-96, hep-th /9603168.}.

We conclude this section by considering the effects of T-duality on
string backgrounds with abelian isometries. The existence of an
abelian isometry
implies that we can choose our coordinates $X^{\mu}=
({\theta}, X^i)$ in such a way
that the spacetime metric is independent of
$\theta$.
This implies that this particular
string solution (string propagating on a target-space admitting an abelian
isometry) is represented by a conformal field theory whose stress
tensor is of the form
$$
T_{g,b}(\sigma)={1\over 2}g^{\theta\theta}(X):\hat{\partial
\theta}\hat{\partial\theta}:(\sigma)+:g^{i\theta}(X)\hat{\partial X_i}
\hat{\partial\theta}:(\sigma)+{1\over 2}:g^{ij}(X)\hat{\partial X_i}
\hat{\partial X_j}:(\sigma).
\numbereq\name{\eqwzopia}
$$

As before, this CFT may be deformed to a point with an enhanced symmetry,
$$
T_{G}(\sigma)={1\over 2}G^{\theta\theta}(X):{\partial{\theta}}{\partial
{\theta}}:(\sigma)+{1\over 2}:G^{ij}(X){\partial X_i}{\partial X_j}:(\sigma)
\numbereq\name{\eqrozap}
$$
and a similar expression for ${\overline T}_{G}(\sigma)$, where
$$
\eqalign{
{\partial{\theta}}&={1\over {\sqrt 2}}({\pi}_{\theta}+G_{\theta\theta}
{\theta}') \cr {\partial X_i}&={1\over {\sqrt 2}}
({\pi}_i+G_{ij}{X^j}') \cr}
\qquad \eqalign{{\overline{\partial}}{\theta}&=
{1\over {\sqrt 2}}({\pi}_{\theta}-G_{\theta\theta}{\theta}') \cr
{\overline{\partial}}X_i&={1\over {\sqrt 2}}({\pi}_i+G_{ij}{X^j}') \cr}
\numbereq\name{\eqsacoer}
$$
The first term in Eq. (\eqrozap) has been defined through a point-splitting
regularization while for the second term we will assume that an adequate
prescription exists in order to define this composite operator so that it
commutes with operators constructed from $\theta$. Note
that Eq.~(\eqrozap) is the direct product of two CFT's which
do not interact with one another---the fields $\theta$ are governed by
a free, toroidal field theory, and they do not interact with the
fields $X^i$, which may have much more general
interactions among themselves.

As should by now be familiar, we must first express $\hat{\partial\theta}$
and $\hat{\partial X_i}$
in terms of ${\partial{\theta}}$,
${\partial X_i}$, ${\overline{\partial}}{\theta}$ and
${\overline{\partial}}X_i$, the light-cone derivatives of the
theory with enhanced symmetry.
$$
\eqalign{
\hat{\partial\theta}&={1\over 2} {\lbrack ({\partial}{\theta}+
{\overline{\partial}}{\theta})+g_{\theta\theta}G^{\theta\theta}(
{\partial}{\theta}-{\overline{\partial}}{\theta})+(g_{\theta i}
+b_{\theta i})G^{ij}({\partial X_j}-{\overline{\partial}}X_j) \rbrack}\cr
\hat{\partial X_i}&={1\over 2} {\lbrack ({\partial}X_i+
{\overline{\partial}}X_i)+(g_{i \theta}+b_{i \theta})G^{\theta\theta}(
{\partial}{\theta}-{\overline{\partial}}{\theta})+g_{i{\kappa}}
G^{{\kappa}j}({\partial X_j}-{\overline{\partial}}X_j)\rbrack} \cr}
\numbereq\name{\eqzopax}
$$
Then the stress-tensor, Eq. (\eqsytu), takes the unnecessarily
intimidating form,
$$
\eqalign{
&T_{g,b}(\sigma)={1\over 8}{\lbrack (g^{\theta\theta}+2G^{\theta\theta}+
g_{\theta\theta}G^{\theta\theta}G^{\theta\theta}+
2g^{\theta i}b_{i \theta}G^{\theta\theta}+g^{ij}b_{i \theta}b_{j \theta}
G^{\theta\theta}){\partial}{\theta}{\partial}{\theta}}+ \cr
&(g^{\theta\theta}-2G^{\theta\theta}+g_{\theta\theta}
G^{\theta\theta}G^{\theta\theta}
-2g^{\theta i}b_{i \theta}G^{\theta\theta}+2g^{ij}b_{i \theta}b_{j \theta}
G^{\theta\theta}){\overline{\partial}}{\theta}
{\overline{\partial}}{\theta}+ \cr
&(2g^{\theta\theta}-2g_{\theta\theta}G^{\theta\theta}G^{\theta\theta}
-g^{ij}b_{i \theta}b_{j \theta}
G^{\theta\theta}){\partial}{\theta}{\overline{\partial}}{\theta}+ \cr
&(2g^{\theta\theta}b_{\theta j}G^{ij}+2g^{\theta i}+2g_{j \theta}G^{\theta
\theta}G^{ij}+2g^{ij}b_{j \theta}G^{\theta\theta} \cr
&\qquad\qquad\qquad\qquad+2g^{\theta j}
b_{j \theta}b_{\theta\kappa}G^{\theta\theta}G^{j \kappa}
+2g^{\kappa j}
b_{j \rho}b_{\kappa \theta}G^{\rho i}G^{\theta \theta})
{\partial}{\theta}{\partial}X_i+ \cr
&(-2g^{\theta\theta}b_{\theta j}G^{ij}+2g^{\theta i}-2g_{j \theta}G^{\theta
\theta}G^{ij}+2g^{ij}b_{j \theta}G^{\theta\theta} \cr
&\qquad\qquad\qquad\qquad-2g^{\theta j}
b_{j \theta}b_{\theta\kappa}G^{\theta\theta}G^{j \kappa}-2g^{\kappa j}
b_{j \rho}b_{\kappa \theta}G^{\rho i}G^{\theta \theta})
{\partial}{\theta}{\overline{\partial}}X_i+ \cr
&(2g^{\theta\theta}b_{\theta j}G^{ij}+2g^{\theta i}-2g_{j \theta}G^{\theta
\theta}G^{ij}-2g^{ij}b_{j \theta}G^{\theta\theta} \cr
&\qquad\qquad\qquad\qquad-2g^{\theta j}
b_{j \theta}b_{\theta\kappa}G^{\theta\theta}G^{j \kappa}-2g^{\kappa j}
b_{j \rho}b_{\kappa \theta}G^{\rho i}G^{\theta \theta})
{\overline{\partial}}{\theta}{\partial}X_i+ \cr
&(-2g^{\theta\theta}b_{\theta j}G^{ij}+2g^{\theta i}+2g_{j \theta}G^{\theta
\theta}G^{ij}-2g^{ij}b_{j \theta}G^{\theta\theta} \cr
&\qquad\qquad\qquad\qquad+2g^{\theta j}
b_{j \theta}b_{\theta\kappa}G^{\theta\theta}G^{j \kappa}+2g^{\kappa j}
b_{j \rho}b_{\kappa \theta}G^{\rho i}G^{\theta \theta})
{\overline{\partial}}{\theta}{\overline{\partial}}X_i+ \cr
&(g^{ij}+g_{\kappa\lambda}G^{\kappa i}G^{\lambda j}+
g^{\theta\theta}b_{\theta\kappa}b_{\theta\lambda}G^{\kappa i}G^{\lambda j}+
g^{\lambda\sigma}b_{\lambda\kappa}b_{\sigma\rho}
G^{\kappa i}G^{\rho j} \cr
&\qquad\qquad\qquad\qquad +2G^{ij}
+2g^{\theta i}b_{\theta\kappa}G^{\kappa j}+
2g^{\lambda j}b_{\lambda\kappa}G^{\kappa i})
{\partial}X_i{\partial}X_j+ \cr
&(g^{ij}+g_{\kappa\lambda}G^{\kappa i}G^{\lambda j}+
g^{\theta\theta}b_{\theta\kappa}b_{\theta\lambda}G^{\kappa i}G^{\lambda j}+
g^{\lambda\sigma}b_{\lambda\kappa}b_{\sigma\rho}
G^{\kappa i}G^{\rho j} \cr
&\qquad\qquad\qquad\qquad
-2G^{ij}-2g^{\theta i}b_{\theta\kappa}G^{\kappa j}-2g^{\lambda j}
b_{\lambda\kappa}G^{\kappa i})
{\overline{\partial}}X_i{\overline{\partial}}X_j+ \cr
&{(g^{ij}-g^{\theta\theta}b_{\theta\kappa}
b_{\theta\lambda}G^{\kappa i}G^{\lambda j}-
g_{\kappa\lambda}G^{\kappa i}G^{\lambda j}
-g^{\lambda\sigma}b_{\lambda\kappa}b_{\sigma\rho}
G^{\kappa i}G^{\rho j}}\cr
&\qquad\qquad\qquad\qquad\qquad {-g^{\lambda j}
b_{\lambda\kappa}G^{\kappa i})
({\partial}X_i{\overline
{\partial}}X_j+{\overline{\partial}}X_i{\partial}X_j) \rbrack}\cr}.
\numbereq\name{\eqwashi}
$$
Since $\theta$ lives on a circle and does not interact with the other
fields, we already know from our previous examples an inner automorphism
that yields a $T$-duality. It is generated by
$$
h={1\over 2i}{\int}d{\sigma}(e^{i{\sqrt 2}{\theta}_L}
-e^{-i{\sqrt 2}{\theta}_L}).
\numbereq\name{\eqropaze}
$$
It's effect on the monstrous Eq. (\eqwashi) is, once again, simply to
change the sign of $\partial \theta$. With a certain amount of tedious
algebra, this transformed stress tensor can be seen to be of the same
form as Eq. (\eqwashi), but with transformed {\it space-time} fields,
{\it i.e.}
$$
e^{i{\pi}h}T_{g,b}(\sigma)e^{-i{\pi}h}=T_{{\tilde g}, {\tilde b}},
\numbereq\name{\eqesaoip}
$$
where,
$$
\eqalign{
{\tilde g}_{\theta\theta}&={1\over g_{\theta\theta}}G^{\theta\theta}
G^{\theta\theta} \cr
{\tilde b}_{\theta i}&={{g_{i \theta}}\over g_{\theta\theta}}
G^{\theta\theta} \cr} \qquad \eqalign{{\tilde g}_{ij}&=g_{ij}-{1\over
{g_{\theta\theta}}}(g_{\theta i}g_{\theta j}-b_{\theta i}b_{\theta j}) \qquad
{\tilde g}_{\theta i}=
-{{b_{\theta i}}\over {g_{\theta\theta}}}G^{\theta\theta}\cr
{\tilde b}_{ij}&=b_{ij}-{1\over {g_{\theta\theta}}}(g_{\theta i}b_{\theta j}
-b_{\theta i}g_{\theta j}). \cr} \numbereq\name{\eqsxeseis}
$$

These transformations were first derived by Buscher, [\buscher]. In
the particular case where $g^{ij}$ is flat, the transformation in
Eq. (\eqsxeseis)
is one of the factorized dualities referred to in our earlier
discussion of the flat torus.

\newsection Duality in spaces without isometries.

In section $3$ we used our general understanding of symmetries to
rederive much of what is currently known about $T$-duality. In so doing,
we hope we persuaded the reader both of the correctness and the
power and simplicity of these techniques. In this section we shall venture
into {\it terra incognita} and demonstrate that our techniques are directly
applicable to general field configurations. We shall show this both through
general arguments and by working out an explicit example. To the
best of our knowledge, no other technique is capable of dealing with
configurations in which the space-time fields are unrepentantly dependent
on the coordinates, $X$.

Recall that conventional techniques can deal quite effectively
with world-sheet
actions that contain only derivatives of $X(\sigma)$. At its
core, $T$-duality simply
interchanges the momentum $\pi(\sigma)$ and $X'(\sigma)$. Unfortunately,
this appears to give a non-local transformation of $X(\sigma)$ itself
$$
X(\sigma)  \to \int^{\sigma} dx \pi(x).
\name\eqdiase
$$
On the other hand our technique implements $T$-duality as an
inner automorphism of the
operator algebra
$$
e^{i{\pi}h}{{\sqrt {\alpha}'}\over R_{cr}}{\pi}(\sigma)e^{-i{\pi}h}=
-{R_{cr}\over {\sqrt {\alpha'}}}X'(\sigma) \qquad e^{i{\pi}h}
{R_{cr}\over {\sqrt {\alpha'}}}X'(\sigma)
e^{-i{\pi}h}=-{{\sqrt {\alpha}'}\over R_{cr}}{\pi}(\sigma).
\numbereq\name{\eqcyira}
$$
For exactly the same reasons, any function of  $X(\sigma)$ transforms in
the same way,
$$
f(X(\sigma)) \to e^{i{\pi}h}f(X(\sigma))
e^{-i{\pi}h}
=f(X(\sigma))
+i{\pi}{\lbrack h, f(X(\sigma)) \rbrack}
+{1\over 2}(i{\pi})^2{\lbrack h, {\lbrack h, f(X(\sigma)) \rbrack}
\rbrack}+\cdots
\numbereq\name{\eqnonlocal}
$$
Recall from section $3$ that the generator that implements $T$-duality is
$$
h={1\over 2i}{\int}d{\sigma}(e^{i{\sqrt 2}X_L}-e^{-i{\sqrt 2}X_L})(\sigma),
\numbereq\name{\eqmweu}
$$
independent of the background fields.
Since any reasonable function may be expanded in Fourier modes, it
is sufficient to consider the effect of $h$ only on such waves. When we
do so, we immediately see that the non-localities of (\eqnonlocal) appear to
manifest themselves in this language also, since
$$
:e^{{\pm i}{\sqrt 2}X_L(z)}::e^{ipX(w, {\overline w})}:=
(z-w)^{{\pm{\sqrt 2}}p}:e^{{\pm i}{\sqrt 2}X_L(z)+ipX(w,
{\overline w})}:
\numbereq\name{\eqabrs}
$$
For general $p$, the RHS of (\eqabrs) has a cut in the complex plane.
When the operator product is converted into a commutator via the usual
deformation of the contour of integration, the discontinuity of
(\eqabrs) across the cut gives rise to a non-local commutator. Alternatively,
it is not hard to show that the commutator of $X_L(\sigma)$ with
itself is non-local,
$$
[ X_L(\sigma), X_L(\sigma')] \sim {\Theta}({\sigma}-{\sigma'})
\numbereq\name{\eqaxoisnm}
$$
leading to similar concerns.

It would appear that nothing short of a miracle can save us, but this is
string theory, and so, of course, a miracle occurs. Observe that the
cut is absent, replaced by a pole when ${\sqrt 2}p$ is an integer. The
absence of
a cut implies, in turn, a purely local commutator. But now recall that we are
compactified on a circle, and so all functions  of $X(\sigma)$ must be
periodic under shifts of $2{\pi}R=2{\sqrt 2}{\pi}$. {\it This periodicity
implies precisely the required quantization condition on
$p={n\over {\sqrt 2}}$.}
(A reminder to the reader: the full operator algebra is the same for all
radii, and we parameterize it using a form natural at $R_{cr}$, hence the
periodicity condition above. This periodicity is the same for all radii;
We have chosen to change
radii by varying {\it metrics}, {\it not\/} by varying periodicities).

Note that the form of equation (\eqabrs) is a consequence of summing an
infinite number of contractions. As such it is an intrinsically
non-perturbative,
world-sheet quantum mechanical result. The miracle occurs only in the
full result, and not in any classical or semi-classical approximation.
(In such approximations, the factor $(z-w)^{-n}$ is replaced by a polynomial
in $\ln (z-w)$, which unavoidably has cuts).

So far, our discussion in this section has been rather general. We therefore
turn now to a simple illustration of these ideas. For the example to be a
sufficiently stringent test of our claims we should consider a solution
corresponding to a space-time possessing curvature, but lacking isometries.
Since curvature is always absent in one dimension, we consider
the bosonic string compactified on a $2$-torus. Thus space-time is
$M^{24} \times T^2$. With a flat metric at the $SU(2)_L^2
\times SU(2)_R^2$ self-dual
point on the torus, the piece of the stress-tensor corresponding to the
torus is
$$
T(\sigma)={1\over 2}{\delta_{\mu\nu}}{\partial}X^{\mu}{\partial}X^{\nu}
\qquad {\overline T}(\sigma)={1\over 2}{\delta_{\mu\nu}}
{\overline\partial}X^{\mu}{\overline\partial}X^{\nu} \qquad
{\mu,\nu=1,2}
\numbereq\name{\eqwesou}
$$
and the two independent duality
transformations are generated by
$$
h_1={1\over 2i}{\int}d{\sigma}(e^{i{\sqrt 2}{X_L}^1}-e^{-i{\sqrt 2}{X_L}^1})
\qquad h_2={1\over 2i}{\int}d{\sigma}(e^{i{\sqrt 2}{X_L}^2}-
e^{-i{\sqrt 2}{X_L}^2})
\numbereq\name{\eqaoyp}
$$
We proceed to turn on curvature by infinitesimally deforming
this particular solution, adding the perturbation
$$
{\delta}T(\sigma)={\delta}{\overline T}(\sigma)=
h_{\mu\nu}(X){\partial}X^{\mu}{\overline\partial}X^{\nu}
\numbereq\name{\eqrorez}
$$
with
$$
h_{\mu\nu}(X)={\epsilon}_{\mu\nu}e^{i{\sqrt 2}X^1}+f_{\mu\nu}
e^{i{\sqrt 2}X^2}+ c. c.
\numbereq\name{\eqwasor}
$$
${\epsilon}_{\mu\nu}$, $f_{\mu\nu}$ are polarization vectors
and $X^{1, 2}={X_L}^{1, 2}+{X_R}^{1, 2}$. On its own, the
deformed field theory is
no longer conformally invariant. However it is straightforward to
restore conformal invariance by making a similar deformation to the
part of the field theory describing the uncompactified $M^{24}$.
The resulting conformal field theory describes the propagation of a massive
Kaluza-Klein excitation in $M^{24}$.
The Riemann tensor
for the curved metric $g_{\mu\nu}={\delta}_{\mu\nu}+h_{\mu\nu}(X)$
in the linearized approximation is given by
$$
R_{1212}=-[{\epsilon}_{22}e^{i{\sqrt 2}X^1}+f_{11}e^{i{\sqrt 2}X^2}+ c.c.]
\numbereq\name{\eqoperat}
$$
In two-dimensions $R_{1212}$ encodes all the information concerning
the curvature of the space. The Ricci scalar is given in terms of
$R_{1212}$
$$
R=2{R_{1212}\over g}=-2[{\epsilon}_{22}e^{i{\sqrt 2}X^1}
+f_{11}e^{i{\sqrt 2}X^2}+ c. c.]
\numbereq\name{\eqpoiuy}
$$
Since the scalar curvature is not constant this two-dimensional curved
space can have at most
one isometry. In the Appendix we demonstrate that the metric
$g_{\mu\nu}(X)={\delta}_{\mu\nu}+h_{\mu\nu}(X)$ actually has no
isometries in general.

To perform a $T$-duality transformation, we need only compute the effect of
a similarity transformation by the operators in equation (\eqaoyp) on the
stress-tensor. There are of course, two factorized dualities, but
for illustrative purposes, we shall consider only the combined duality which
is their product.
$$
\eqalign{
e^{i{\pi}h_2}e^{i{\pi}h_1}(T+{\delta}T)(\sigma)e^{-i{\pi}h_1}e^{-i{\pi}h_2}
&={1\over 2}{\delta}_{\mu\nu}e^{i{\pi}h_2}e^{i{\pi}h_1}
:{\partial}X^{\mu}{\partial}X^{\nu}:e^{-i{\pi}h_1}e^{-i{\pi}h_2} \cr
&+e^{i{\pi}h_2}e^{i{\pi}h_1}:h_{\mu\nu}(X){\partial}X^{\mu}
{\overline\partial}X^{\nu}:e^{-i{\pi}h_1}e^{-i{\pi}h_2} \cr}
\numbereq\name{\eqorder}
$$
Since $h_{\mu\nu}(X)$ is given by Eq. (\eqwasor) we must define what we
mean by normal-ordered operators in the expression above
$$
:e^{i{\sqrt 2}X_L}{\partial}X(\sigma):={\lim_{{\epsilon}\to 0}} {\lbrack
:e^{i{\sqrt 2}X_L}(\sigma):{\partial}X({\sigma}+{\epsilon})
-{2i:e^{i{\sqrt 2}X_L(\sigma)}:\over 4{\pi}{\epsilon}} \rbrack}
\numbereq\name{\eqashlp}
$$
and a similar expression for the anti-holomorphic part which commutes
with the holomorphic one in the normal-ordered expression.
In order to derive the effect of the inner automorphism on the
stress-tensor we need to know how the automorphisms act on
${\partial}X^{\mu}$, ${\overline\partial}X^{\mu}$, $e^{i{\sqrt 2}X^{\mu}_L}$
and $e^{i{\sqrt 2}X^{\mu}_R}$. The effect on the first two is known
from the previous section
$$
e^{i{\pi}h}{\partial X}^{\mu}(\sigma)
e^{-i{\pi}h}=-{\partial X}^{\mu}(\sigma) \qquad e^{i{\pi}h}
{\overline{\partial}}X^{\mu}(\sigma)e^{-i{\pi}h}
={\overline{\partial}}X^{\mu}(\sigma)
\numbereq\name{\eqaoue}
$$
while the effect on the last two can be computed by writing
$$
\eqalign{
e^{i{\pi}h_1}e^{i{\sqrt 2}(X^{1}_L+X^{1}_R)(\sigma)}e^{-i{\pi}h_1}&=
e^{i{\sqrt 2}X^{1}_R}(\sigma) (e^{i{\sqrt 2}X^{1}_L}(\sigma)+i{\pi}[h_1,
e^{i{\sqrt 2}X^{1}_L}(\sigma)] \cr
&+{1\over 2}(i{\pi})^2[h_1, [h_1,
e^{i{\sqrt 2}X^{1}_L}(\sigma)]]+ \cdots ) \cr}
\numbereq\name{\eqasoput}
$$
and computing the commutators. This can be achieved by going
from the cylinder to the complex plane, calculating the corresponding
operator product expansions and computing the residue at the pole.
The result then is given by
$$
e^{i{\pi}h_1}e^{i{\sqrt 2}(X^{1}_L+X^{1}_R)(\sigma)}e^{-i{\pi}h_1}=
-e^{-i{\sqrt 2}(X^{1}_L-X^{1}_R)}(\sigma)
\numbereq\name{\eqegw}
$$
With Equations (\eqaoue) and (\eqegw) at our disposal we can now calculate
the effect of the inner automorphism (duality transformation) on the
stress-tensor. This reads
$$
\eqalign{
&e^{i{\pi}h_2}e^{i{\pi}h_1}(T+{\delta}T)(\sigma)e^{-i{\pi}h_1}e^{-i{\pi}h_2}
\cr &={1\over 2}{\delta}_{\mu\nu}e^{i{\pi}h_2}e^{i{\pi}h_1}
{\lim_{{\epsilon}\to 0}}[{\delta}_{\mu\nu}
{\partial}X^{\mu}(\sigma){\partial}X^{\nu}({\sigma}+{\epsilon})
+{1\over 2{\pi}{\epsilon}^2}]e^{-i{\pi}h_1}e^{-i{\pi}h_2} \cr
&+e^{i{\pi}h_1}e^{i{\pi}h_2}{\lim_{{\epsilon}\to 0}}[{\epsilon}_{\mu\nu}
:e^{i{\sqrt 2}X^{1}_L}:(\sigma){\partial}X^{\mu}(\sigma+\epsilon)\cr
&\qquad\qquad\qquad-{2i:e^{i{\sqrt 2}X^{1}_L(\sigma)}:\over 4{\pi}{\epsilon}}]
:e^{i{\sqrt 2}X^{1}_R}{\overline\partial}X^{\nu}:(\sigma)
e^{-i{\pi}h_1}e^{-i{\pi}h_2} \cr
&+e^{i{\pi}h_1}e^{i{\pi}h_2}{\lim_{{\epsilon}\to 0}}[f_{\mu\nu}
:e^{i{\sqrt 2}X^{2}_L}:(\sigma){\partial}X^{\mu}(\sigma+\epsilon)\cr
&\qquad\qquad\qquad-{2i:e^{i{\sqrt 2}X^{2}_L(\sigma)}:\over 4{\pi}{\epsilon}}]
:e^{i{\sqrt 2}X^{2}_R}{\overline\partial}X^{\nu}:(\sigma)
e^{-i{\pi}h_1}e^{-i{\pi}h_2} \cr
&={1\over 2}{\delta}_{\mu\nu}:{\partial}X^{\mu}{\partial}X^{\nu}:
+:({\epsilon}_{\mu\nu}e^{-i{\sqrt 2}(X^{1}_L-X^{1}_R)}+f_{\mu\nu}
e^{-i{\sqrt 2}(X^{2}_L-X^{2}_R)}){\partial}X^{\mu}{\overline\partial}X^{\nu}:
\cr}
\numbereq\name{\eqniko}
$$
Equation (\eqniko) is our result. Where the starting stress-tensor described a
massive Kaluza-Klein excitation in $M^{24}$, the transformed stress-tensor
in equation (\eqniko) describes a winding excitation. The physical
interpretation therefore is both transparent and unsurprising.
We observe that up to a sign the result of the duality transformation is to
replace the $X_{+}(\sigma)=X_L(\sigma)+X_R(\sigma)$ dependence of the
metric with its dual coordinate $X_{-}=X_L(\sigma)-X_R(\sigma)$.
On its own, the dual coordinate $ X_{-} $ is a very sick non-local operator,
but it is a marvelous and well-known fact that all the operators in (\eqniko) are local.
The deformation of the original conformal field theory corresponds to turning
on a Kaluza-Klein graviton $h_{\mu\nu}(X)$, a momentum excitation of the
string spectrum. The effect of T-duality is to transform the KK graviton into
a winding mode excitation and subsequently the two conformal field theories
which result from perturbing the original theory by sending weak gravitational
KK waves and weak winding waves should be identified. In the following
paragraphs we will
derive similar results for arbitrary field configurations on tori.

Two more comments on equation (\eqniko) are appropriate. First, the reader
may be wondering what happened to $R \to {1\over R}$. The answer
is that we considered a non-constant deformation of the theory at the
critical radius. $T$-Duality turned the metric excitation into a winding
excitation, but inverting the metric produced no visible effect precisely
because we started at the critical radius. We could, however, have
started just as easily at any radius; the constant piece of the metric
would transform just as it did in section $3$, and the constant piece of the
metric would have been inverted. Indeed, each Fourier mode transforms
independently as a simple consequence of the linearity of the
similarity transformation ; we chose not to repeat the calculation of
section $3$, solely to avoid cluttering our formulas.

Our second comment relates to the size of our non-constant deformation.
In the above analysis we chose to consider weak
perturbations; ${\epsilon}_{\mu\nu}$
and $f_{\mu\nu}$ were taken to be infinitesimal. We did so
only because this makes the analysis of the conditions for conformal invariance
straightforward [\ovre]. However, these conditions actually have
no effect on the
form of the $T$-duality transformation. Conformal field theories may
be constructed
by writing down general translation invariant {\it ans\"{a}tze\/} for
$T(\sigma)$
and ${\overline T}(\sigma)$, and then demanding that they
satisfy Virasoro $\times$ Virasoro [\eaw], which
imposes conditions (called equation of
motion!) on the space-time fields. These are Popeye conditions---they are
what they are---but they do not affect in any way symmetry transformations
on those fields; they just come along for the ride. ( Since imposing
Virasoro $\times$ Virasoro is an {\it algebraic} condition, it is
of course preserved by
similarity transformations). Thus our analysis of $T$-duality transformations
is immediately extendable to finite deformations.

So far we have discussed space-time fields which are constant or which
have the lowest mode on the torus excited. We end this
section with a derivation
of the transformation of higher modes. The first thing we need to understand
is the transformation of pure functions of $X$; that is, we need to
compute $e^{i{\pi}h}e^{i{\sqrt 2}n(X_L+X_R)(\sigma)}e^{-i{\pi}h}$,
where $n \in {\BbbZ}$. With this result, we shall then be able to compute the
transformation of arbitrary periodic functions of $X(\sigma)$ multiplied by
arbitrary light-cone derivatives of $X(\sigma)$ by the same
point-splitting techniques
we used in (\eqniko). To compute this quantity we could in principle use
the same technique we used above to compute the transformation of the
lowest mode, expanding the exponential in powers of $h$ and computing
the multiple commutators. However, for the higher mode such an
approach becomes increasingly more involved as $n$ increases and,
with it, the order of poles from contractions. Rather, it is easier to
use the result for $n=1$ and induction. First, we write
$$
: e^{i{\sqrt 2}nX_L(\sigma)}:={\lim_{{\epsilon} \to 0}}
:e^{i{\sqrt 2}X_L(\sigma+\epsilon)}::e^{i{\sqrt 2}(n-1)X_L(\sigma)}:
({1\over {\epsilon}})^{(n-1)\over {\pi}}
\numbereq\name{\eqgoezu}
$$
It then follows that
$$
\eqalign{
e^{i{\pi}h}:e^{i{\sqrt 2}nX_L(\sigma)}:&e^{-i{\pi}h} \cr
&=
{\lim_{{\epsilon} \to 0}}e^{i{\pi}h}
:e^{i{\sqrt 2}X_L(\sigma+\epsilon)}:e^{-i{\pi}h}
e^{i{\pi}h}:e^{i{\sqrt 2}(n-1)X_L(\sigma)}:e^{-i{\pi}h}
({1\over {\epsilon}})^{(n-1)\over {\pi}} \cr}
\numbereq\name{\eqasopiue}
$$
and we may use equation (\eqegw)
and induction to compute the result for arbitrary $n$.
The result is
$$
e^{i{\pi}h}:e^{i{\sqrt 2}nX_L(\sigma)}:e^{-i{\pi}h}=(-1)^n
:e^{-i{\sqrt 2}nX_L(\sigma)}:
\numbereq\name{\eqease}
$$
Since $X_L(\sigma)$ and $X_R(\sigma)$ are constructed from mutually
commuting sets of creation and annihilation operators it follows that
$$
e^{i{\pi}h}:e^{i{\sqrt 2}n(X_L+X_R)(\sigma)}:e^{-i{\pi}h}=(-1)^n
:e^{-i{\sqrt 2}n(X_L-X_R)(\sigma)}:
\numbereq\name{\eqzwiax}
$$
Thus, modulo additional terms coming from contraction with light-cone
derivatives, the $n$-th momentum mode is interchanged with the
$n$-th winding mode.

We now have all the pieces we need in order to compute the
transformation on any field. A general term in the stress tensor
will be of the, ``weighted tensor," type [\asios],
$$
{\phi_{\mu\nu\cdots\rho}}(X){\partial^{w_1}}X^{\mu}{\partial^{w_2}}
X^{\nu} \cdots {\partial^{w_n}}X^{\kappa}
{\overline{\partial}}^{v_1}X^{\lambda} \cdots {\overline{\partial}}^{v_m}
X^{\rho}
 \numbereq\name{\eqkorad}
$$
Again, we may decompose such a term into Fourier modes, and
consider the transformation of each separately under $T$-Duality. Using
equations (\eqzwiax) and (\eqaoue), we
might expect the transformation to be,
$$
\eqalign{
e^{i{\sqrt 2}p(X_L+X_R)}&{\partial^{w_1}}X^{\mu}{\partial^{w_2}}
X^{\nu} \cdots {\partial^{w_n}}X^{\kappa}
{\overline{\partial}}^{v_1}
X^{\lambda} \cdots {\overline{\partial}}^{v_m}X^{\rho}(\sigma) \to \cr
& \to (-1)^{n+p}
e^{-i{\sqrt 2}p(X_L-X_R)(\sigma)}{\partial^{w_1}}X^{\mu}{\partial^{w_2}}
X^{\nu} \cdots {\partial^{w_n}}X^{\kappa}
{\overline{\partial}}^{v_1}
X^{\lambda} \cdots {\overline{\partial}}^{v_m}X^{\rho}(\sigma) \cr}
\numbereq\name{\eqgenmodetransf}
$$
This result is, in fact, correct, but it is not quite trivial to
demonstrate it because of the normal-ordering present in
such terms. However, the point-splitting method we have used
repeatedly throughout this paper is applicable in this
general case, and an inductive proof of equation (\eqgenmodetransf) is
not hard to construct. We shall not give the details here (they are
straightforward, if a little tedious), but the first step is to prove the
result for terms involving any mode and a single $n$'th derivative
of $X$, and then to inductively increase the number of derivatives
of $X$. This result implies that the transformation on {\it any\/} field
$\phi$, with $n$ indices contracting with holomorphic derivatives
of $X$ is simply
$$
\phi (X_+) \longleftrightarrow (-1)^n \phi ({-X_-} + \pi/\sqrt{2}).
\numbereq\name\eqgenftransf
$$
Equation (\eqgenftransf) gives us the transformation of
arbitrary space-time fields, and is our final result.

\newsection Discussion.

What have we done in this paper? We have taken a general approach
to understanding the symmetries of string theory and applied it
to $T$-duality. We have rederived known results when background
fields are constant (or can be made so), and we have shown how a string
miracle enables us to deal with general field configurations on
space-time tori. We worked out an explicit example of such a case, and
derived a simple general formula for the transformation of
arbitrary background fields. In our opinion, $T$-duality on tori is
now fully understood (or, at least, understandable).

What have we not done? We have worked only on tori, and
have ignored other topologies. Unfortunately, at our present level of
understanding, each space-time topology must be studied
separately, at least when
studying $T$-duality. The reason is that the algebra of observables associated
with a conformal field theory, while the same for all CFTs in a deformation
class, appears to change when we change to another deformation
class ({\it i.e.} another space-time topology). How does $T$-duality
manifest itself on other topologies? Is there a relationship between
$T$-duality and mirror symmetry
\ref{\miro}{L.\tie Dixon in {\it Superstrings, Unified Theories and
Cosmology}, G. Furlan et al. eds., (World Scientific, Singapore
1988); W.\tie Lerche, C.\tie Vafa and N.\tie Warner,
\np324 (1989), 427;
B. R.\tie Greene and M. R.\tie Plesser, \np338
(1990), 15.}? Does some avatar of $T$-duality remain to
influence string theory in uncompactified space-time? Does
there exist some Ur-theory that would enable us to handle
all space-time topologies simultaneously?

Similarly, we may wonder about the relationship of the present work
to the currently fashionable $S$-duality. At first sight, our techniques
would appear to be inapplicable to $S$-duality. After all, while our
methods are fully quantum-mechanical on the world sheet, we have
worked only on $S^2$, {\it i.e.} only at string tree-level. $S$-duality, being
a strong-weak coupling duality would appear to require a fully quantum
mechanical understanding of string theory. However, recent very
interesting work \ref{\ram}{M. J.\tie Duff, \np442 (1995), 47;
M. J.\tie Duff and R. R.\tie Khuri, \np411 (1994) 473,
M. J.\tie Duff, J. T.\tie Liu and
J. \tie Rahmfeld, CTP-TAMU-27/95 preprint, hep-th/9508094;
J. \tie Schwarz and
A. \tie Sen, \np411 (1994), 35.}
seems to suggest an intimate connection
between $S$ and $T$-dualities, and to do so at an entirely
classical level of analysis. Perhaps the techniques of this paper will
be able to throw some light on on $S$-duality after all.

\newsection Acknowledgments.

We would like to thank I. Bakas, E. Kiritsis and J. Liu for useful discussions.
This work was supported in part by the Department of Energy Contract
Number DE-FG02-91ER40651-TASKB. 

\immediate\closeout1
\bigbreak\bigskip

\line{\bf References.\hfil}
\nobreak\medskip\vskip\parskip

\input refs

\newsection Appendix.

In this appendix we shall demonstrate that the metric $g_{\mu\nu}(X)
={\delta}_{\mu\nu}+h_{\mu\nu}(X)$ with
$$
h_{\mu\nu}(X)={\epsilon}_{\mu\nu}e^{i{\sqrt 2}X^1}+f_{\mu\nu}
e^{i{\sqrt 2}X^2}+ c. c.
\numbereq\name{\eqway}
$$
does not possess any isometries. The problem of determining
all infinitesimal isometries of the metric $g_{\mu\nu}(X)$ is
reduced to the problem of determining all Killing vectors of the metric.
The linearized Ricci scalar for $g_{\mu\nu}(X)$ is given by
$$
R={\partial_{\alpha}}{\partial^{\alpha}}h^{\mu}_{\mu}-{\partial^{\mu}}
{\partial^{\nu}}h_{\mu\nu}=(-p^2{\epsilon}^{\mu}_{\mu}+p^{\mu}p^{\nu}
{\epsilon}_{\mu\nu})e^{ipX}+(-q^2f^{\mu}_{\mu}+q^{\mu}q^{\nu}f_{\mu\nu})
e^{iqX}+ c.c
\numbereq\name{\eqapee}
$$
with $p^{\mu}={\sqrt 2}(1, 0)$ and $q^{\mu}={\sqrt 2}(0, 1)$. If $V^{\mu}$ is a
Killing vector then $V^{\mu}{\partial_{\mu}}R=0$. Since $R$ is
not constant then $V^\mu$ must be tangent to curves
of constant $R$. Thus there
can be at most one Killing vector field. A solution to the relation
$V^{\mu}{\partial_{\mu}}R=0$ is provided by
$$
V^\lambda=p^\lambda(-q^2f^{\mu}_{\mu}+q^{\mu}q^{\nu}f_{\mu\nu})
e^{iqX}-q^\lambda(-p^2{\epsilon}^{\mu}_{\mu}+p^{\mu}p^{\nu}
{\epsilon}_{\mu\nu})e^{ipX}+ c.c
\numbereq\name{\eqasnop}
$$
By finding the curves of constant $R$ one may characterize possible
Killing vectors up to a scalar function $\phi(X)$. It suffices then to
prove that there is no non-trivial scalar function $\phi(X)$ that satisfies
as a result of the Killings equation the relation
$$
{\partial_{(\kappa}}(e^{\phi}V_{\lambda)})=e^{\phi}
(V_{\lambda}{\partial_{\kappa}}
{\phi}+V_{\kappa}{\partial_{\lambda}}
{\phi}+{\partial_{\kappa}}V_{\lambda}+{\partial_{\lambda}}V_{\kappa})=0
\numbereq\name{\eqpoxc}
$$
Contracting the indices of (\eqpoxc) with $p^{\lambda}, p^{\kappa}$
($q^{\lambda}, q^{\kappa}$) and taking into account the form of $V^\mu$
(\eqasnop) we find
$$
p^\mu{\partial_{\mu}}{\phi}=0 \qquad  q^\mu{\partial_{\mu}}{\phi}=0
\numbereq\name{\eqroipu}
$$
The only solution to the above equations is that of a constant function
which fails to yield a solution to Killing's equation for generic
polarization vectors $f_{\mu\nu}$, $\epsilon_{\mu\nu}$. Thus
the metric $g_{\mu\nu}(X)$
in general possess no isometries.

\vfill\end

\bye